\documentclass[twocolumn,twoside,showpacs,superscriptaddress,nofootinbib]{revtex4}

\usepackage[dvips]{graphicx}
\usepackage[dvips]{color}
\usepackage{epsfig}
\usepackage{amssymb,amsmath}
\usepackage{lscape}

\RequirePackage{xspace}

\newcommand{\dkpp}{\ensuremath{D\to K^0_S\pi^+\pi^-}\xspace}
\newcommand{\dnkpp}{\ensuremath{D^0\to K^0_S\pi^+\pi^-}\xspace}
\newcommand{\dbkpp}{\ensuremath{\overline{D}{}^0\to K^0_S\pi^+\pi^-}\xspace}
\newcommand{\bdkp}{\ensuremath{B^0\to D K^+\pi^-}\xspace}
\newcommand{\bdnkp}{\ensuremath{B^0\to D^0 K^+\pi^-}\xspace}
\newcommand{\bdbkp}{\ensuremath{B^0\to \overline{D}{}^0 K^+\pi^-}\xspace}
\newcommand{\bdcpkp}{\ensuremath{B^0\to D_{CP} K^+\pi^-}\xspace}
\newcommand{\bdkstar}{\ensuremath{B^0\to D K^{*0}}\xspace}

\newcommand{\bdk}{\ensuremath{B^+\to D K^+}\xspace}
\newcommand{\bdcpk}{\ensuremath{B^+\to D_{CP} K^+}\xspace}

\newcommand{\dkp}{\ensuremath{D\to K^+\pi^-}\xspace}
\newcommand{\dkpbar}{\ensuremath{D\to K^-\pi^+}\xspace}

\newcommand{\ab}{\ensuremath{A_B}\xspace}
\newcommand{\abbar}{\ensuremath{\overline{A}_B}\xspace}
\newcommand{\bvar}{\ensuremath{(m^2_{D\pi}, m^2_{K\pi})}\xspace}

\newcommand{\ad}{\ensuremath{A_D}\xspace}
\newcommand{\adbar}{\ensuremath{\overline{A}_D}\xspace}
\newcommand{\dvar}{\ensuremath{(m^2_{+}, m^2_{-})}\xspace}

\newcommand{\aab}{\ensuremath{|A_B|}\xspace}
\newcommand{\aabbar}{\ensuremath{|\overline{A}_B|}\xspace}
\newcommand{\aad}{\ensuremath{|A_D|}\xspace}
\newcommand{\aadbar}{\ensuremath{|\overline{A}_D|}\xspace}

\begin{document}

\title{Double Dalitz Plot Analysis of the Decay $ B^0 \to D K^+ \pi^- $, $ D \to K^0_S \pi^+ \pi^- $}

\author{Tim Gershon}
\affiliation{Department of Physics, University of Warwick, Coventry CV4 7AL, United Kingdom}
\author{Anton Poluektov}
\affiliation{Department of Physics, University of Warwick, Coventry CV4 7AL, United Kingdom}
\affiliation{Budker Institute of Nuclear Physics, 11 Lavrentieva, Novosibirsk, 630090, Russia}

\date{\today}

\begin{abstract}
  It is shown that it is possible to perform a model-independent extraction 
  of the CKM Unitarity Triangle angle $\gamma$ using only the decays 
  \bdkp with \dkpp and \bdkp with flavour-specific $D$ decays. 
  The proposed method can also utilise the \bdkp data with $CP$-eigenstate
  decays of the $D$ meson in a model-independent way.
\end{abstract}


\maketitle

\section{Introduction}

Among the fundamental parameters of the Standard Model of particle physics,
the angle $\gamma = {\rm arg}\left(-V_{ud} V^*_{ub}/V_{cd}V^*_{cb}\right)$ of 
the Unitarity Triangle formed from elements of the Cabibbo-Kobayashi-Maskawa 
quark mixing matrix~\cite{Cabibbo:1963yz,Kobayashi:1973fv}
has a particular importance.
Not only is it one of the least well determined fundamental parameters of the
Standard Model, it is also the only $CP$-violating parameter that can be
measured using only tree-level decays~\cite{Antonelli:2009ws}.
The precise determination of $\gamma$ is thus a critical element of the
programme towards understanding the baryon asymmetry of the Universe,
and is one of the main objectives of planned future $B$ physics experiments 
(see, for example,~\cite{Browder:2007gg,Buchalla:2008jp,Browder:2008em}).

A method to measure the CKM phase $\gamma$ in \bdkp decays by comparing the
Dalitz plot distribution obtained when the neutral $D$ meson is reconstructed
in $CP$ eigenstates to that when flavour-specific states are used has been
proposed recently~\cite{Gershon:2008pe}.
The method builds on the original proposal of Gronau, London and Wyler
(GLW)~\cite{Gronau:1990ra,Gronau:1991dp} for a similar analysis using 
\bdk decays, and exploits the fact that $CP$ violation effects are expected to
be enhanced in $B^0 \to DK^{*0}$ decays since the interfering amplitudes are of
comparable magnitude~\cite{Bigi:1988ym}.  Since the charge of the kaon in
\bdkp decays unambiguously tags the flavour of the decaying $B$ meson, there
is no need for time-dependent analysis~\cite{Dunietz:1991yd}.

Compared to previous proposals for similar
analyses~\cite{Gronau:2002mu,Pruvot:2007yd,Aleksan:2002mh} this method has
several attractive features: 1) the amplitudes can be measured directly in
the Dalitz plot analysis, relative to the flavour-specific $B^0 \to
D_2^{*-}K^+$ amplitude; 2) there is no need to normalise to decays with
another final state; 3) there is sufficient information to extract $\gamma$
using only $D$ decays to $CP$-even states;\footnote{
  $CP$-odd $D$ decays with large branching fractions 
  such as $D \to K_S\pi^0$ are difficult to reconstruct,
  particularly in experiments at hadron colliders.
} 4) the sensitivity to $\gamma$ does
not depend strongly on the values of $\gamma$ and strong phases involved; 5)
$\gamma$ is determined with only a single unresolvable ambiguity ($\gamma
\longrightarrow \gamma + \pi$).
These advantages have much in common with those of the \bdk, \dkpp method,
which is currently providing the most constraining measurements of $\gamma$~\cite{Aubert:2008bd,Abe:2008wya}.
Moreover, this technique is especially promising for the LHCb experiment since
it does not require the reconstruction of neutral particles in the final state.

There is, however, a potential drawback of the method, in that the fit to the
Dalitz plot distribution is inherently model-dependent.  The magnitude of the
model-dependence has been estimated to be small~\cite{Gershon:2009qc}, but in
view of the aim to achieve an ${\cal O}(1^\circ)$ measurement of $\gamma$ at
future experiments, it is certainly desirable to explore possibilities to
remove such systematic uncertainties.

A model-independent analysis procedure has been proposed for the extraction of
$\gamma$ from \bdk, \dkpp decays~\cite{Giri:2003ty,Bondar:2008hh}.  
The method uses data from charm factory experiments -- specifically,
$CP$-tagged charm mesons -- to obtain information
about the strong phase variation across the $D$ decay Dalitz plot.
A similar approach to probe the strong phases present in the \bdkp Dalitz plot
is not immediately available since there is no realistic way to obtain
sufficient samples of $CP$-tagged $B$ decays. 

In this paper we show that the study of Dalitz plot distributions in the decay
chain \bdkp, \dkpp allows to translate the model-independent measurement of
the \dkpp amplitude to the \bdkp mode.  
This approach allows not only to measure $\gamma$ model-independently in this
decay chain, but also to use the obtained constraints on the \bdkp amplitude 
in the \bdcpkp Dalitz plot analysis.  
For clarity, our method is independent of model assumptions in both \bdkp and
\dkpp decays.  The more straightforward cases which are either model-dependent
in \bdkp but model-independent in \dkpp, or vice-versa, are also possible but
are not discussed in detail.  A fully model-dependent analysis is, of course,
also possible.

The remainder of the paper is organised as follows.  
In Section~\ref{sec:idea} we outline the basic idea and the formalism of the
analysis.  
In Section~\ref{sec:Dflav} we discuss how possible complications arising from
the use of $D\to K\pi$ decays as quasi-flavour-specific states can be
resolved.
In Section~\ref{sec:models} we describe the Dalitz plot models that we use in
our feasibility study.
We finally present the results of the study in Section~\ref{sec:sensitivity},
before concluding in Section~\ref{sec:conclusion}. 

\section{Formalism}
\label{sec:idea}

We begin by recalling the essentials of the \bdk, \dkpp model-independent
method~\cite{Giri:2003ty,Bondar:2008hh}.
The amplitude of the \bdk, \dkpp decay can be written
\begin{equation}
  A_{D\,{\rm Dlz}}=\adbar+ r_Be^{i(\delta_B+\gamma)}\ad\,, 
\end{equation}
where $\adbar=\adbar(m^2_{K_S\pi^+}, m^2_{K_S\pi^-})\equiv\adbar\dvar$ 
is the amplitude of the \dbkpp decay,  $\ad=\ad\dvar$ 
is the amplitude of the \dnkpp decay, 
$r_B$ is the ratio of the absolute values of the interfering 
$B^+\to \overline{D}{}^0K^+$ and $B^+\to D^0K^+$ amplitudes, 
and $\delta_B$ is the strong phase difference between these amplitudes. 
Assuming no $CP$ violation in $D$ decay, $\ad\dvar = \adbar(m^2_{-},m^2_{+})$. 
The density of the $D$ decay Dalitz plot from \bdk\ decay is given by the 
absolute value squared of the amplitude 
\begin{eqnarray}
  |A_{D\,{\rm Dlz}}|^2 & = & |\adbar+ r_Be^{i(\delta_B+\gamma)}\ad|^2 =   
  \label{p_b} \\
  & &\aadbar^2+r_B^2\aad^2+2\aad\aadbar(xc-ys)\,, \nonumber
\end{eqnarray}
where 
\begin{equation}
  x = r_B\cos(\delta_B+\gamma)\,; \;\;\;
  y = r_B\sin(\delta_B+\gamma)\,.
\end{equation}
The functions $c=c\dvar$ and $s=s\dvar$ are the cosine and sine of the strong 
phase difference $\delta_D=\arg\ad-\arg\adbar$ between the \dnkpp and \dbkpp 
amplitudes\footnote{
  This paper follows the convention for strong phases in $D$ decay
  amplitudes used by CLEO~\cite{Briere:2009aa}. 
  Note that alternative conventions are used in the literature, 
  {\it e.g.} Ref.~\cite{Bondar:2008hh} uses a convention where the 
  sign of $\delta_D$ is opposite.
}: 
\begin{equation}
  c=\cos\delta_D(m^2_+,m^2_-)\,; \;\;\;
  s=\sin\delta_D(m^2_+,m^2_-)\,. 
\end{equation}
The equations for the charge-conjugate mode $B^-\to D K^-$ are obtained 
with the substitution $\gamma \longrightarrow -\gamma$. Considering both 
$B$ charges, one can obtain $\gamma$ and $\delta_B$ separately. 

Once the Dalitz plot is divided into $2\mathcal{N}$ bins symmetrically
to the exchange $m^2_-\leftrightarrow m^2_+$, 
the expected number of events  in the $i^{\rm th}$ bin of the Dalitz plot of
$D$ decay from \bdk\ is  
\begin{equation}
  \langle N_i\rangle = 
  h_{D\,{\rm Dlz}} \left[
    K_i + r_B^2K_{-i} + 2\sqrt{K_iK_{-i}}(xc_i-ys_i)
  \right] \,, 
  \label{n_b}
\end{equation}
where $K_i$ is the number of events in the corresponding bin of the Dalitz plot 
where the $D$ meson is in a flavour eigenstate (conveniently obtained using
$D^{*\pm}\to D\pi^\pm$ samples) and $h_{D\,{\rm Dlz}}$ is a normalisation constant. 
The bin index $i$ ranges from $-\mathcal{N}$ to $\mathcal{N}$ (excluding 0); 
the exchange $m^2_+ \leftrightarrow m^2_-$ corresponds to the exchange 
$i\leftrightarrow -i$. 
The coefficients $c_i$ and $s_i$, which include information about 
the cosine and sine of the phase difference, are given by
\begin{equation}
  c_i=\frac{\int\limits_{\mathcal{D}_i}
            \aad\aadbar
            \cos\delta_D\,d\mathcal{D}
            }{\sqrt{
            \int\limits_{\mathcal{D}_i}\aad^2 d\mathcal{D}
            \int\limits_{\mathcal{D}_i}\aadbar^2 d\mathcal{D}
            }}\,, 
  \label{cs}
\end{equation}
where the dependences on Dalitz plot position of $\aad$, $\aadbar$ and
$\delta_D$ have been suppressed.
The parameters $s_i$ are defined similarly with cosine substituted by sine. 
Here $\mathcal{D}$ represents the Dalitz plot phase space and 
$\mathcal{D}_i$ is the bin region over which the integration is performed. 

The symmetry under $\pi^+ \leftrightarrow \pi^-$ requires $c_i = c_{-i}$ and
$s_i = - s_{-i}$, 
thus for $2\mathcal{N}$ bins there are $2\mathcal{N}+3$ unknowns.
For each bin there are two observables in $B\to DK$ decays, being the 
numbers of events in $B^+$ and $B^-$ samples,
and thus there are a total of $4\mathcal{N}$ observables.  
The system can consequently be solved for $\mathcal{N} \ge 2$.
However, due to the small value of $r_B$ in \bdk decays, there is very little
sensitivity to the $c_i$ and $s_i$ parameters, leading to a reduction in the
precision on $\gamma$ that can be obtained.
Use of external constraints on $c_i$ and $s_i$, that can be provided by
charm-factory experiments~\cite{Briere:2009aa}, leads to much improved
sensitivity to $\gamma$.

Turning now to \bdkp decays, we find that the simplest approach towards
performing a model-independent analysis is not viable. 
Using the convention $D_{CP} = \left( D^0 \pm \bar{D}^0 \right) / \sqrt{2}$,
where $+$ and $-$ correspond to $CP$-even and $CP$-odd final states,
respectively,
the amplitude for \bdcpkp decay is given by
\begin{equation}
  A_{B\,{\rm Dlz}} = 
  \frac{1}{\sqrt{2}} \left( \pm \abbar + e^{i\gamma} \ab \right)\,, 
\end{equation}
where $\abbar=\abbar\bvar$ is the amplitude of the \bdbkp decay and
$\ab=\ab\bvar$ is the amplitude of the \bdnkp decay.
In these expressions we have factored out the
$CP$-violating phase $\gamma$ so that the amplitudes \abbar and \ab
are $CP$-conserving. Both the strong phase difference 
between \bdbkp and \bdnkp decays and the ratio of the suppressed and
allowed amplitudes $r_B$ are now functions of $B$ decay Dalitz plot variables
and enter the expressions for amplitudes.

Let us now consider dividing the $B$ decay Dalitz plot into
bins (denoted by the index $\alpha$, $1\le \alpha\le \mathcal{M}$).
The number of expected events in each \bdcpkp Dalitz plot bin is
\begin{equation}
  \begin{array}{rcl}
    \langle M_{\alpha} \rangle & = & 
    h_{B\,{\rm Dlz}} \Big[ \overline{N}_{\alpha} + N_{\alpha} \pm  \\
    & & \hspace{11mm}
    2 \sqrt{N_{\alpha}\overline{N}_{\alpha}}
    (\varkappa_{\alpha}\cos\gamma - \sigma_{\alpha}\sin\gamma) \Big] \, .
  \end{array}
  \label{num_bin_cp}
\end{equation}
In this expression the factors $\overline{N}_{\alpha}$ and $N_{\alpha}$
give the numbers of events in the corresponding bin in \bdbkp and \bdnkp
decays, respectively (complications arising due to the difficulty to
reconstruct $D$ mesons in pure flavour eigenstates are discussed in
Section~\ref{sec:Dflav} below).
The factor $h_{B\,{\rm Dlz}}$ is a global normalisation constant.
In Eq.~(\ref{num_bin_cp}) we have introduced the parameters $\varkappa_\alpha$
and $\sigma_\alpha$ that describe the average strong phase difference and
amplitude difference between \bdbkp and \bdnkp decays in each bin, and that
are defined by: 
\begin{equation}
  \varkappa_{\alpha}=
  \frac{
    \int\limits_{\mathcal{D}_{\alpha}}
    |\ab||\abbar|\cos\delta_B\,d\mathcal{D}
  }{
    \sqrt{
      \int\limits_{\mathcal{D}_{\alpha}}|\ab|^2 d\mathcal{D}
      \int\limits_{\mathcal{D}_{\alpha}}|\abbar|^2 d\mathcal{D}
    }
  }\,, 
  \label{kappasigma}
\end{equation}
where the dependence on the $B$ decay Dalitz plot position of $|\ab|$,
$|\abbar|$ and $\delta_B=\arg\ab-\arg\abbar$ 
have been suppressed,
$\mathcal{D}$ again represents the phase space (now of the $B$ decay Dalitz
plot) and $\mathcal{D}_\alpha$ is the bin region over which the integration is
performed.  
The parameters $\sigma_{\alpha}$ are given by similar expressions with cosine
substituted by sine.

Until this stage the discussion for \bdkp decays has run in parallel to that
for \bdk, \dkpp.  However, unlike the \bdk, \dkpp case, the two amplitudes
$\abbar$ and $\ab$ are inherently different and thus there is no symmetry in
the Dalitz plot to be exploited.  
In the case of $CP$-eigenstate $D$ decays, we therefore have $2\mathcal{M}+1$ 
unknowns (even if the normalisation is fixed externally), that cannot be 
solved for with $2\mathcal{M}$ observables.
Hence external constraints on $\varkappa_\alpha$ and $\sigma_\alpha$ are
necessary~\cite{Gronau:2002mu}.

Let us now consider both Dalitz plots simultaneously in the \bdkp, \dkpp 
decay chain. The amplitude can be written as
\begin{equation}
  A_{\rm DblDlz}=\abbar\adbar + e^{i\gamma} \ab\ad\,. 
\end{equation}
The decay probability is proportional to the absolute value squared 
of the amplitude $A$: 
\begin{equation}
  \begin{array}{rcl}
    |A_{\rm DblDlz}|^2 & = & \aabbar^2\aadbar^2 + \aab^2\aad^2 \, + \\
    \multicolumn{3}{l}{
      \hspace{12mm}
      2 \aab\aad\aabbar\aadbar \, \times
    } \\
    \multicolumn{3}{l}{
      \hspace{24mm}
      {\rm Re} \left( e^{i\gamma} e^{i\delta_B\bvar}e^{i\delta_D\dvar} \right)
    } \\
    & = & \aabbar^2\aadbar^2 + \aab^2\aad^2 \, + \\
    \multicolumn{3}{l}{
      \hspace{12mm}
      2\aab\aad\aabbar\aadbar \, \times
    } \\
    \multicolumn{3}{l}{
      \hspace{24mm}
      \left[
        (\varkappa c - \sigma s)\cos\gamma -
        (\varkappa s + \sigma c)\sin\gamma 
      \right] \, .
    }
  \end{array}
  \label{ampl_sq}
\end{equation}
To implement the model-independent analysis both $B$ and $D$ plots have to be
binned.  Corresponding bin indices are denoted here, as before, by Greek and
Latin characters, respectively. The expected number of events in the bins of
\bdkp, \dkpp double Dalitz plot is then
\begin{equation}
  \begin{array}{rcl}
    \langle M_{\alpha i} \rangle = & 
    h_{\rm DblDlz} \Big\{ \overline{N}_{\alpha}K_{i} + N_{\alpha}K_{-i} \, + \\
    \multicolumn{3}{l}{
      \hspace{8mm}
      2 \sqrt{N_{\alpha}K_{i}\overline{N}_{\alpha}K_{-i}} \, \times 
    } \\
    \multicolumn{3}{l}{
      \hspace{16mm}
      \left[
        (\varkappa_{\alpha}c_i - \sigma_{\alpha}s_i)\cos\gamma - 
        (\varkappa_{\alpha}s_i + \sigma_{\alpha}c_i)\sin\gamma 
      \right] \Big\} \, ,
    }
  \end{array}
  \label{num_bin}
\end{equation}
where the definition of the event numbers $N_{\alpha}$, $\overline{N}_{\alpha}$, 
$K_i$ and phase terms $c_i, s_i, \varkappa_{\alpha}, \sigma_{\alpha}$ is the same 
as for the processes mentioned above, and $h_{\rm DblDlz}$ is a normalisation
constant.

If the number of bins in the \bdkp Dalitz plot is $\mathcal{M}$ and the number
of bins in the \dkpp Dalitz plot is $2\mathcal{N}$, then after exploiting the
symmetry of the \dkpp decay as before, the number of equations 
represented by Eq.~(\ref{num_bin}) is $\mathcal{MN}$.
The number of unknowns (including the normalisation factor $h_{\rm DblDlz}$) is
$2\mathcal{M}+2\mathcal{N}+2$.
Naturally, information on $\gamma$ cannot be extracted from the decays
of only one $B$ flavour, as is apparent since Eq.~(\ref{num_bin}) is invariant
under the rotation of $\gamma$ with the simultaneous rotation of all
$(\varkappa,\sigma)$ pairs. 
To determine the $CP$ violation parameter it is, of course, necessary to add
the opposite $B$ flavour (with the substitution $\gamma\longrightarrow -\gamma$).  
The system can now be fully resolved: 
in total, there are $2\mathcal{MN}$ observables and 
$2\mathcal{M}+2\mathcal{N}+2$ unknowns, 
{\it i.e.} it is solvable for $(M-1)(N-1) \ge 2$.
Since the coefficients $c_i$ and $s_i$ are the same as in the
model-independent $B\to DK$ analysis, and can be obtained at a charm factory,
the number of unknowns can be reduced even further.

As a by-product of the \bdkp, \dkpp analysis, the coefficients
$\varkappa_{\alpha}$ and $\sigma_{\alpha}$ will be determined.
The values of these parameters can then be used to enable a model-independent
analysis of \bdcpkp decays. Technically the combination of \bdkp, \dkpp and
\bdcpkp modes is conveniently done by using a combined likelihood fit that
includes both the expressions Eq.~(\ref{num_bin}) and Eq.~(\ref{num_bin_cp}). 

\section{\boldmath Effects of Doubly-Cabibbo-Suppressed Contributions to $D\to K\pi$ decays}
\label{sec:Dflav}

In practise it is hard to reconstruct secondary $D$ mesons from $B$ decays in
pure flavour eigenstates.  (The possibility to exploit the charge of the
associated pion in $D^{*\pm}\to D\pi^\pm$ is not available.)
Semileptonic $D^0$ decays in principle provide a source of pure
flavour-specific states, but owing to small branching fractions and large
backgrounds these are experimentally difficult to deal with,
especially at hadron experiments such as LHCb.  It is more convenient to use
hadronic final states such as $K\pi$, which are not pure flavour tags due to
the presence of a small admixture of doubly-Cabibbo-suppressed amplitude.
This contribution can significantly affect a precision measurement. 

In the proposed analysis, hadronic final states can be used after some
correction of the procedure.
The amplitude of a \bdkp, \dkp decay is 
\begin{equation}
  A_{\rm fav} = \abbar + r_{K\pi} e^{-i\delta_{K\pi}} e^{i\gamma} \ab \, ,
\end{equation}
where $r_{K\pi}$ and $\delta_{K\pi}$ are the ratio of the magnitudes of the
suppressed and favoured $D$ decay amplitudes and the strong phase between
them, respectively.\footnote{
We use a sign convention for $\delta_{K\pi}$ that is consistent with
that used in the majority of the literature (see, {\it e.g.}
Ref.~\cite{Antonelli:2009ws}), but is opposite in sign to that used for
$\delta_D$.  Another different convention is used by
CLEO~\cite{Rosner:2008fq,Asner:2008ft}.
}
The number of events in the \bdkp Dalitz plot bins with $D$ detected 
in a $K^+\pi^-$ state is: 
\begin{equation}
  \begin{array}{rcl}
    \langle N^{\rm fav}_{\alpha} \rangle & = & 
    \overline{N}_{\alpha} + r_{K\pi}^2 N_{\alpha} + 
    2r_{K\pi} \sqrt{N_{\alpha}\overline{N}_{\alpha}} \, \times \\
    & & \hspace{3mm} \left[
      \varkappa_{\alpha}\cos(-\delta_{K\pi}+\gamma)-
      \sigma_{\alpha}\sin(-\delta_{K\pi}+\gamma)
    \right]\,. 
    \label{bin_num_kp}
  \end{array}
\end{equation}

Similarly for the suppressed decay \bdkp, \dkpbar, the amplitude is 
\begin{equation}
  A_{\rm sup} = r_{K\pi} e^{-i\delta_{K\pi}} \abbar + e^{i\gamma} \ab \, , 
\end{equation}
and the number of events is 
\begin{equation}
  \begin{array}{rcl}
    \langle N^{\rm sup}_{\alpha} \rangle & = & 
    r_{K\pi}^2\overline{N}_{\alpha} + N_{\alpha} + 
    2r_{K\pi} \sqrt{N_{\alpha}\overline{N}_{\alpha}} \, \times \\
    & & \hspace{3mm} \left[
      \varkappa_{\alpha}\cos(\delta_{K\pi}+\gamma)-
      \sigma_{\alpha}\sin(\delta_{K\pi}+\gamma)
    \right]\,. 
    \label{bin_num_kpbar}
  \end{array}
\end{equation}
The expressions for $\overline{B}{}^0\to DK^-\pi^+$ decays are obtained 
by the substitution $\gamma\to -\gamma$. 

The relations of Eq.~(\ref{bin_num_kp}) and Eq.~(\ref{bin_num_kpbar}) add
another  $4\mathcal{M}$ equations to the system of equations to be used in the
analysis ($2\mathcal{M}$ equations for each $B$ flavour), but since the
constants $r_{K\pi}$ and $\delta_{K\pi}$ are known from charm
analyses~\cite{Rosner:2008fq,Asner:2008ft}, only $2\mathcal{M}$ new unknowns
are added (the numbers $N_{\alpha}$ and $\overline{N}_{\alpha}$).
Hence, the inclusion of the suppressed final states into the analysis can help
to improve the overall sensitivity to
$\gamma$~\cite{Atwood:1996ci,Atwood:2000ck}.  Recent studies suggest further
improvement could be gained from including also the $D \to K\pi\pi^0$
decays~\cite{Lowery:2009id}.

\section{\boldmath $B$ Decay Dalitz Plot Models}
\label{sec:models}

In order to perform a feasibility study of the proposed method it is necessary
to define a $B$ decay Dalitz plot model. 
The sensitivity depends crucially on the structure of the \bdnkp and \bdbkp
amplitudes, specifically on the interference between them. 
If the strong phase difference $\delta_B$ is nearly constant ({\it e.g.} in
the case when both amplitudes are dominated by $B\to DK^*$ decay), the double
Dalitz analysis reduces to a conventional Dalitz plot analysis technique
similar that used for  $B^{\pm}\to DK^{\pm}$ decays. 
Note that in this type of analysis the possible contribution from non-$K^*$
states can be taken into account model-independently by introducing the
coherence factor $0\leq \kappa\leq 1$ to the interference
term~\cite{Gronau:2002mu} (analyses using this approach have been performed,
albeit with low statistics~\cite{Aubert:2008yn,:2009au}).
However, if the resonance structure in \bdkp decays is rich with multiple
overlapping states in different channels ($K^*$, $D^*$ and $D_s^*$), the
proposed technique should provide a significant benefit compared to analysis
of the \dkpp\ Dalitz plot with the coherence factor. 

\begin{table*}
  \caption{Parameters of \bdkp amplitudes. The suppressed \bdnkp amplitude 
           uses $r_B=0.4$}
  \label{ampl_table}
  \begin{tabular}{|l||c|c|c|c|c|c|}
  \hline\hline
       & \multicolumn{3}{c|}{\bdbkp}  & \multicolumn{3}{c|}{\bdnkp} \\
       \cline{2-7}
             & Amplitude & Phase & Fit frac.& Amplitude & Phase & Fit frac. \\ \hline
 $K^*(892)$     & $1$    & $0^{\circ}$    & $47$\% & $r_B\times 1$     & $120^{\circ}$ & $56$\%\\
 $K^*_0(1430)$  & $0.2$  & $284^{\circ}$  & $0.015$\%& $r_B\times 0.2$ & $284^{\circ}$ & $0.018$\%\\
 $K^*_2(1430)$  & $0.16$ & $221^{\circ}$  & $10$\% & $r_B\times 0.16$  & $221^{\circ}$ & $12$\%\\
 $K^*(1680)$    & $0.61$ & $128^{\circ}$  & $2.4$\%& $r_B\times 0.61$  & $128^{\circ}$ & $2.9$\%\\
 $D^*_0(2400)$  & $4.8$  & $267^{\circ}$  & $2.3$\%& -                 & -             & \\
 $D^*_2(2460)$  & $1.1$  & $325^{\circ}$  & $30$\% & -                 & -             & \\
 $D_{sJ}(2700)$ & -      & -              &        & $r_B\times 4.0$   & $200^{\circ}$ & $25$\%\\
 Non-res        & $1.25$ & $140^{\circ}$  & $5.3$\%& $r_B\times 1.25$  & $140^{\circ}$ & $6.3$\%\\
  \hline\hline
  \end{tabular}
\end{table*}

Unlike the \dkpp amplitude, the amplitudes for \bdkp decays are not yet well 
established. 
For our feasibility study, based on pseudo-experiments generated using Monte
Carlo (MC) simulation, we use several models based on two-body amplitudes with
$K^*$ and $D^*$ resonances. 
We define first model A, which includes only $K^*(892)$ and $D^*_2(2460)$
states in the favoured amplitude and $K^*(892)$ in the suppressed amplitude
with relative magnitudes consistent with measured values for the branching
fractions~\cite{Aubert:2005yt} and phases chosen arbitrarily.
Model B is similar to that used in a recent study~\cite{Gershon:2009qc}, where
the contributions of further resonances are based on results from $B\to
D^+K^0\pi^-$~\cite{Aubert:2007qe} using the assumption of isotopic symmetry. 
Model C includes a contribution from the $D_{sJ}(2700)$
resonance~\cite{:2007aa} in addition to all states of model B. 
Its contribution is estimated from the ratio 
\begin{equation}
  \frac{\mathcal{B}(B\to D_{sJ}\pi)}{\mathcal{B}(B\to D_{sJ}D)}\simeq
  \frac{\mathcal{B}(B\to D_{s}^{(*)}\pi)}{\mathcal{B}(B\to D_{s}^{(*)}D)}
\end{equation}
with a phase space correction. 
Models D and E use the same states as models B and C, respectively,
excluding the nonresonant term.
For the \dkpp amplitude, we use the model from the most recent Belle
analysis~\cite{:2008wya}.

\begin{figure*}
  \centering
  \epsfig{file=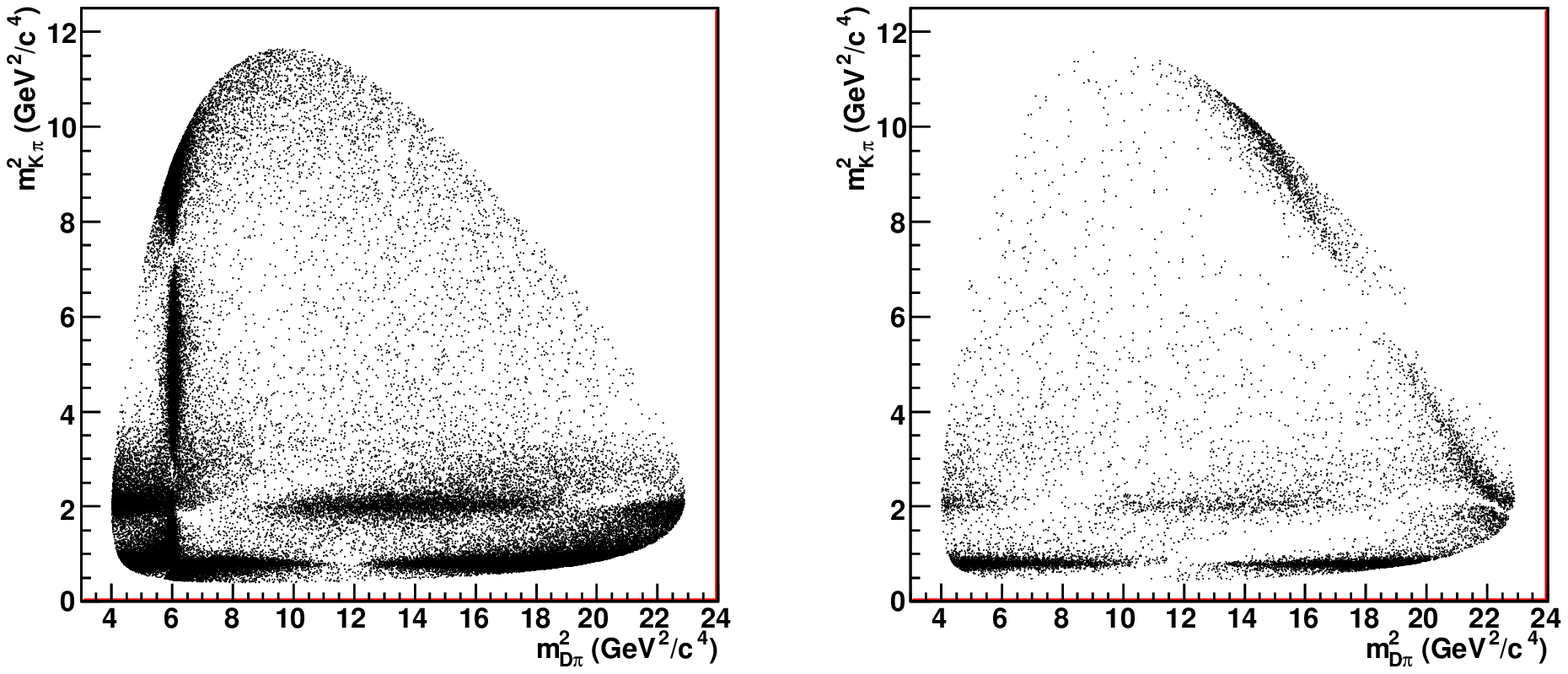,width=0.8\textwidth}
  \put(-250,150){\mbox{\bf(a)}}
  \put(-45,150){\mbox{\bf(b)}}
  \caption{Generated Dalitz plots for (a) \bdbkp and (b) \bdnkp decays
  using amplitude model C.}
  \label{bdkp_plots}
\end{figure*}

The fit fractions and phases for model C are shown in Table~\ref{ampl_table}. 
Other models contain subsets of the states from model C with the same
amplitudes and phases.
The Dalitz plots of \bdbkp\ and \bdnkp\ decays generated according to model C 
are shown in Figure~\ref{bdkp_plots}. 

\begin{figure*}
  \centering
  \epsfig{file=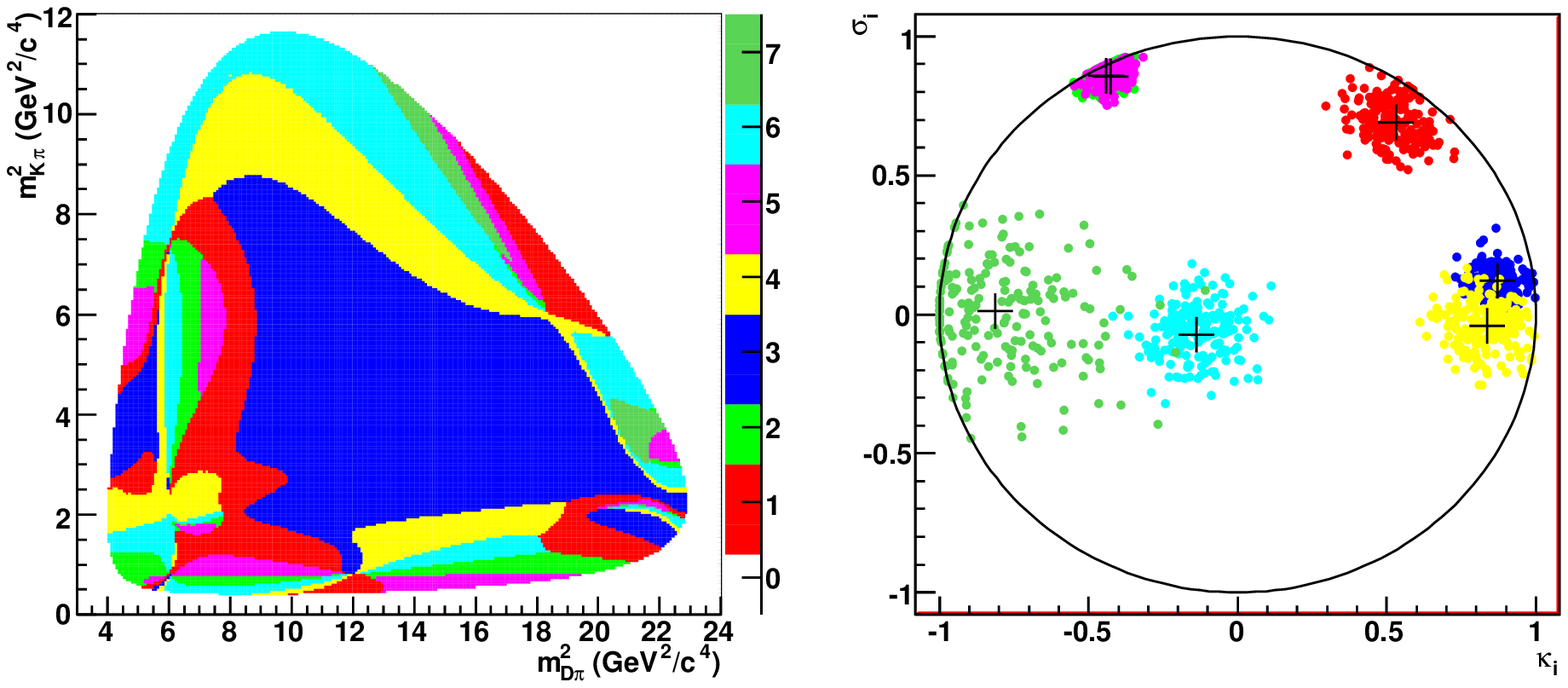,width=0.8\textwidth}
  \put(-250,150){\mbox{\bf(a)}}
  \put(-45,150){\mbox{\bf(b)}}
  \caption{(a) Binning of \bdkp Dalitz plot and (b) values of $\varkappa_i$, 
           $\sigma_i$ parameters for optimal binning (crosses are calculated 
           values, scattered points are the values obtained from the fit
           to toy MC samples).}
  \label{bdkp_bins}
\end{figure*}

\begin{figure*}
  \centering
  \epsfig{file=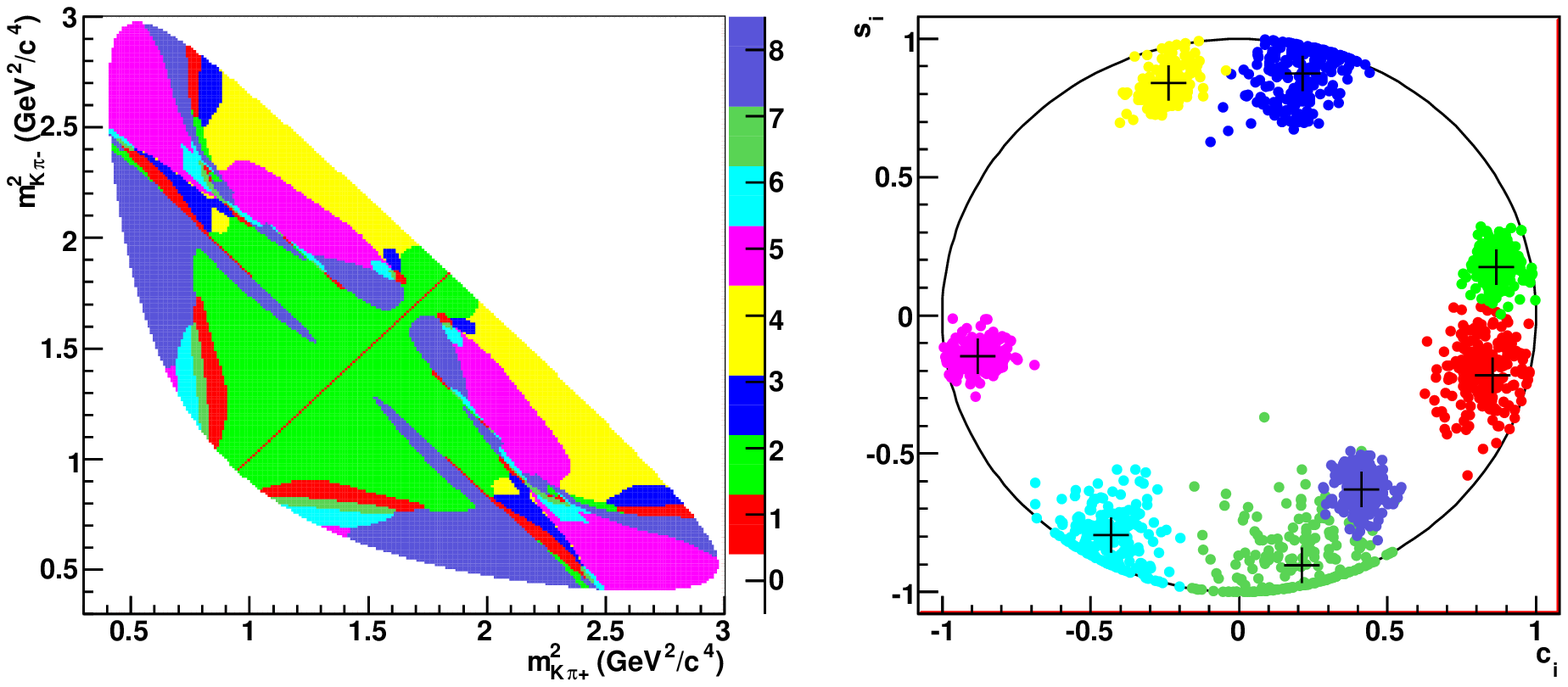,width=0.8\textwidth}
  \put(-250,150){\mbox{\bf(a)}}
  \put(-45,150){\mbox{\bf(b)}}
  \caption{(a) Binning of \dkpp Dalitz plot and (b) values of $c_i$, 
           $s_i$ parameters for optimal binning (crosses are calculated 
           values, scattered points are the values obtained from the fit
           to toy MC samples).}
  \label{dkpp_bins}
\end{figure*}

The sensitivity of the technique will depend on the choice of binning for 
both \dkpp and \bdkp Dalitz plots. The choice of optimal binning for \dkpp
Dalitz plot has been previously considered in application to \bdk
modes~\cite{Bondar:2008hh}.
Here the requirements for the optimal binning are the same: the strong 
phase difference ($\delta_B$ in case of \bdkp plot and $\delta_D$ for \dkpp 
plot) and the ratio of suppressed to allowed amplitudes should be as 
constant as possible over each bin to maximise values of 
$\varkappa_{\alpha}^2+\sigma_{\alpha}^2$ and $c_i^2+s_i^2$. 
In the case of the \dkpp Dalitz plot, a binning based on uniform division of
strong phase difference has been found to give a good approximation to the
optimal~\cite{Bondar:2008hh}.  
However, in the case of \bdkp amplitudes, such an approach gives poor results
because of very large variations of the absolute value of amplitude ratio
across the bin.  A binning based on maximising the ``binning quality factor''
$Q$~\cite{Bondar:2008hh}, gives significantly better results. 
The optimal binning of the \bdkp Dalitz plot with 7 bins for model C
is shown in Figure~\ref{bdkp_bins} (a), 
while Figure~\ref{bdkp_bins} (b) shows the values of 
$\left( \varkappa_{\alpha}, \sigma_{\alpha} \right)$ obtained in each bin.
For the \dkpp Dalitz plot, we also use optimised binning obtained 
with the same technique (see Figure~\ref{dkpp_bins}) with 
binning quality factor Q=0.89.

We use the values $r_B=0.4$ and $\gamma = 60^{\circ}$ in the simulation. 
Note that the value of $r_B$ can differ for different $K^*$ states 
({\it e.g.} $K^*(892)$, $K^*_0(1430)$, $K^*_2(1430)$ {\it etc.}); we use the
same value for all these states.  
We generate pseudo-experiments with samples of $10^4$ events for each $B$
flavour for both \bdkp with \dkpp and \bdcpkp modes. 
Note that once the Dalitz plot models are fixed, the ratios of the 
numbers of events between the different samples is also fixed.
The numbers of \bdbkp and \bdnkp decays are $10^5$ and $\sim 8000$,
respectively. These numbers of events roughly correspond to 20 times the
expected annual yield of the LHCb experiment.\footnote{
  A full simulation of the detector response to \bdkp, \dkpp decays has not
  been performed, but we assume that the ratio of the efficiency of this mode
  to \bdcpkp mode is equal to that between \bdk, \dkpp and \bdcpk decays, 
  where roughly equal annual yields are expected~\cite{Akiba:2008zz}.
}
Hence the precision obtained from this study is indicative of that which could
be achieved in an upgraded phase of LHCb operation.  It is, however, trivial
to scale the sensitivities to lower luminosities.

\section{Results of the Feasibility Study}
\label{sec:sensitivity}

\begin{table*}
  \caption{Sensitivity of $\gamma$ measurement with different techniques
           using \bdkp decays}
  \label{sens_table}
  \begin{tabular}{rlccccc}\hline\hline
      &                          & Model A & Model B & Model C 
                                 & Model D & Model E\\
    \hline
      & Composition              & \begin{tabular}{c} $K^*$, \\ $D^*_2$ only \end{tabular}
                                 & \begin{tabular}{c} No $D_{sJ}$ \\ \end{tabular}
                                 & \begin{tabular}{c} All states \\ included \end{tabular}
                                 & \begin{tabular}{c} No $D_{sJ}$ and \\ no nonres. \end{tabular}
                                 & \begin{tabular}{c} No nonres. \\ \end{tabular}
                                 \\
    \hline                          
      & \bdkp binning quality $Q$& 0.90  & 0.91 & 0.81 & 0.86 & 0.77 \\
    \hline
    1 & \begin{tabular}{l} Unbinned fit of \bdcpkp  \end{tabular} 
                                 & $2.47^{\circ}\pm 0.16^{\circ}$
                                 & $1.31^{\circ}\pm 0.07^{\circ}$ 
                                 & $1.17^{\circ}\pm 0.06^{\circ}$
                                 & $1.71^{\circ}\pm 0.09^{\circ}$
                                 & $1.53^{\circ}\pm 0.08^{\circ}$\\
    \hline
    2 & \begin{tabular}{l} Binned fit of \bdcpkp \\ (fixed $\varkappa, \sigma$) \end{tabular}
                                 & $3.45^{\circ}\pm 0.33^{\circ}$ 
                                 & $1.56^{\circ}\pm 0.08^{\circ}$ 
                                 & $1.46^{\circ}\pm 0.07^{\circ}$ 
                                 & $2.07^{\circ}\pm 0.11^{\circ}$
                                 & $1.86^{\circ}\pm 0.10^{\circ}$\\
    \hline
    3 & \begin{tabular}{l} Unbinned Dalitz fit of \bdkstar, \\ \dkpp \end{tabular}
                                 & $1.34^{\circ}\pm 0.07^{\circ}$ 
                                 & $1.50^{\circ}\pm 0.08^{\circ}$ 
                                 & $1.54^{\circ}\pm 0.08^{\circ}$
                                 & $1.46^{\circ}\pm 0.08^{\circ}$
                                 & $1.50^{\circ}\pm 0.08^{\circ}$\\
    \hline
    4 & \begin{tabular}{l} Binned Dalitz fit of \bdkstar, \\ \dkpp \end{tabular}
                                 & $1.48^{\circ}\pm 0.07^{\circ}$
                                 & $1.71^{\circ}\pm 0.09^{\circ}$ 
                                 & $1.73^{\circ}\pm 0.09^{\circ}$
                                 & $1.72^{\circ}\pm 0.09^{\circ}$
                                 & $1.78^{\circ}\pm 0.09^{\circ}$\\
    \hline
    5 & \begin{tabular}{l} Binned \bdkstar, \dkpp fit \\
      (floated $c_i$, $s_i$) \end{tabular}
                                 & $2.03^{\circ}\pm 0.11^{\circ}$ 
                                 & $2.67^{\circ}\pm 0.18^{\circ}$ 
                                 & $2.54^{\circ}\pm 0.16^{\circ}$
                                 & $2.40^{\circ}\pm 0.15^{\circ}$
                                 & $2.40^{\circ}\pm 0.15^{\circ}$\\
    \hline
    6 & \begin{tabular}{l} Binned double Dalitz fit\\
      (floated $\varkappa, \sigma$) \end{tabular}
                                 & $1.37^{\circ}\pm 0.09^{\circ}$ 
                                 & $1.49^{\circ}\pm 0.08^{\circ}$ 
                                 & $1.45^{\circ}\pm 0.07^{\circ}$
                                 & $1.46^{\circ}\pm 0.07^{\circ}$
                                 & $1.50^{\circ}\pm 0.08^{\circ}$\\
    \hline
    7 & \begin{tabular}{l} Binned double Dalitz fit \\
      (floated $\varkappa, \sigma$, $c_i$, $s_i$) \end{tabular}
                                 & $1.81^{\circ}\pm 0.09^{\circ}$ 
                                 & $1.56^{\circ}\pm 0.08^{\circ}$ 
                                 & $1.51^{\circ}\pm 0.08^{\circ}$
                                 & $1.51^{\circ}\pm 0.08^{\circ}$
                                 & $1.59^{\circ}\pm 0.08^{\circ}$\\
    \hline
    8 & \begin{tabular}{l} Binned double Dalitz fit with \\ \bdcpkp 
      (floated $\varkappa, \sigma$) \end{tabular}
                                 & $1.29^{\circ}\pm 0.06^{\circ}$ 
                                 & $1.21^{\circ}\pm 0.06^{\circ}$ 
                                 & $1.20^{\circ}\pm 0.06^{\circ}$
                                 & $1.39^{\circ}\pm 0.07^{\circ}$
                                 & $1.31^{\circ}\pm 0.07^{\circ}$\\
    \hline
    9 &  \begin{tabular}{l} Binned double Dalitz fit with \\ \bdcpkp 
      (floated $\varkappa, \sigma$, $c_i$, $s_i$) \end{tabular} 
                                 & $1.52^{\circ}\pm 0.08^{\circ}$ 
                                 & $1.27^{\circ}\pm 0.06^{\circ}$ 
                                 & $1.24^{\circ}\pm 0.06^{\circ}$
                                 & $1.36^{\circ}\pm 0.07^{\circ}$
                                 & $1.41^{\circ}\pm 0.07^{\circ}$\\
    \hline\hline
  \end{tabular}
\end{table*}

Different strategies for using \bdkp decays to measure $\gamma$ are studied
using MC simulation. 
The results are shown in Table~\ref{sens_table}. 
The uncertainties are obtained from the spread of results in the samples of
200 pseudo-experiments.

An unbinned fit of \bdcpkp decays with free parameters $r_B$, $\gamma$
and $\delta_B$ (option 1) gives an estimate of the statistical error for a
model-dependent Dalitz plot analysis of this decay~\cite{Gershon:2008pe}.
A binned fit of the same mode (option 2) with fixed values of $\varkappa$ and
$\sigma$ shows a drop of sensitivity due to binning of the \bdkp Dalitz plot. 
For most amplitude models, the loss in sensitivity is about 20\%. 

We next consider selecting only events in the $K^*$ region of \bdkp phase
space, introducing a coherence factor~\cite{Gronau:2002mu}, and using \dkpp
decays.
A conventional single Dalitz plot analysis (option 3) gives comparable
sensitivity to the unbinned \bdcpkp analysis (option 1).
However, only a fraction of all \bdkp events are used indicating that further
gains are potentially possible. 
Changing from unbinned to a binned model-independent approach (option 4)
reduces the sensitivity by about $15$\%.
Relaxing the $c_i$ and $s_i$ coefficients in the binned approach is also
possible (option 5)~\cite{Giri:2003ty}. 
Although the sensitivity in this case reduces significantly, the reduction is
far less than in the $B^{\pm}\to DK^{\pm}$ mode~\cite{Bondar:2005ki} due to
the larger interference term in neutral $B$ decays and also due to the optimal
binning used in this study.

The double Dalitz plot analysis (option 6) uses all available events 
in the \bdkp phase space, providing better $\gamma$ accuracy compared to
option 4, and also giving as a by-product the values of \bdkp amplitude
coefficients $\varkappa_{\alpha}$ and $\sigma_{\alpha}$. 
The values of $\varkappa_{\alpha}$ and $\sigma_{\alpha}$ returned from the fit
and a comparison with their calculated values is shown in
Figure~\ref{bdkp_bins} (b).
Relaxing the \dkpp amplitude coefficients $c_i$, $s_i$ (option 7) 
results in a much smaller reduction of the sensitivity than that with single 
Dalitz plot analysis selecting only the $K^*$ region. 
The fitted values of the $c_i$, $s_i$
coefficients for this option are shown in Figure~\ref{dkpp_bins} (b). 

Combined with \bdcpkp decays (option 8), the double Dalitz analysis allows 
to further improve the statistical precision. 
However the gain in sensitivity for most models is about $15\%$ worse than the
weighted error for options 6 and 2, which means that the precision of
$\varkappa_{\alpha}$ and $\sigma_{\alpha}$ terms obtained from the Dalitz
analysis affects the \bdcpkp mode.
With very large statistics it may be possible to mitigate this effect by using
a finer binning of the \bdkp Dalitz plot. 

\begin{figure*}
  \centering
  \epsfig{file=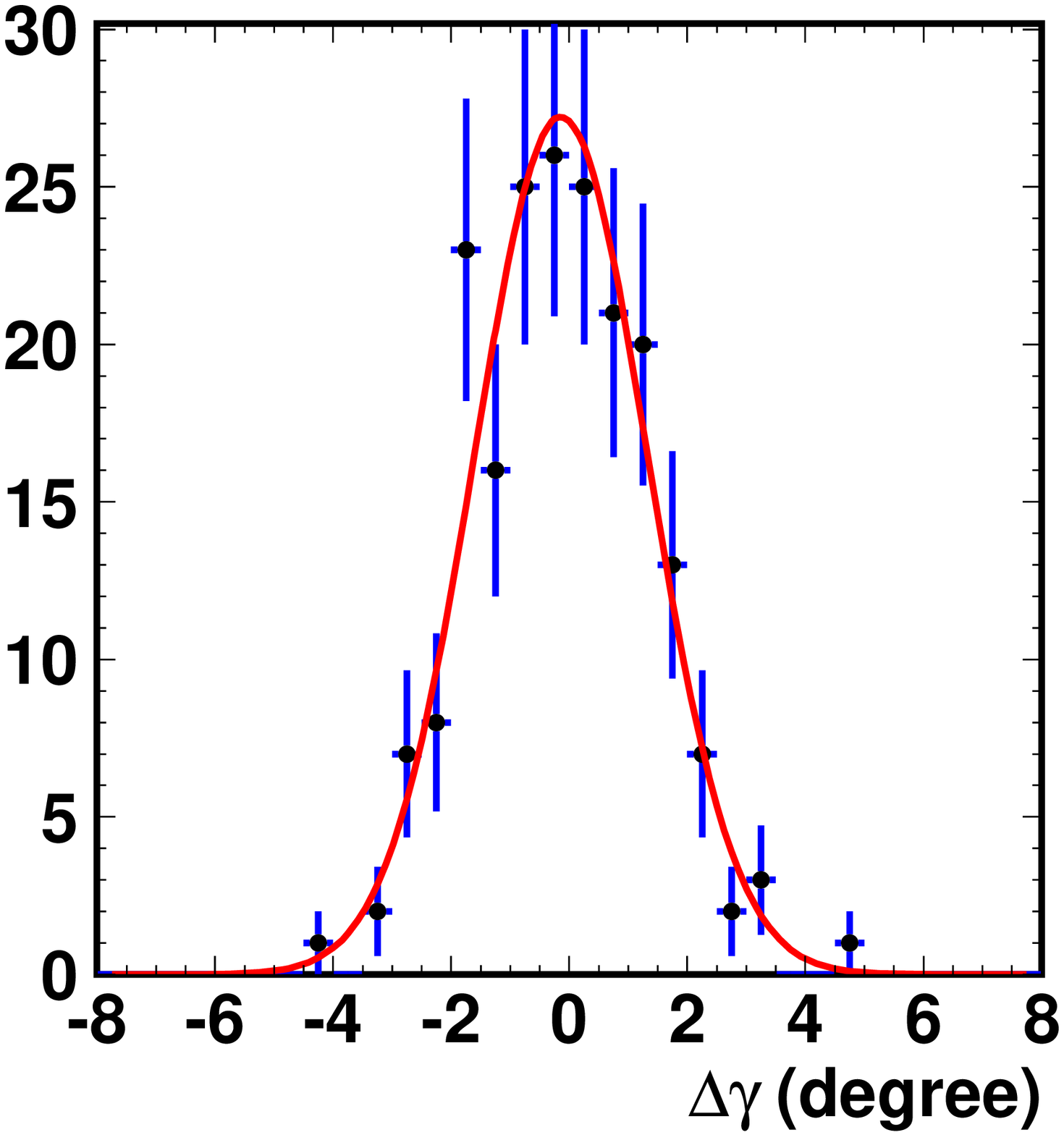, width=0.3\textwidth}
  \epsfig{file=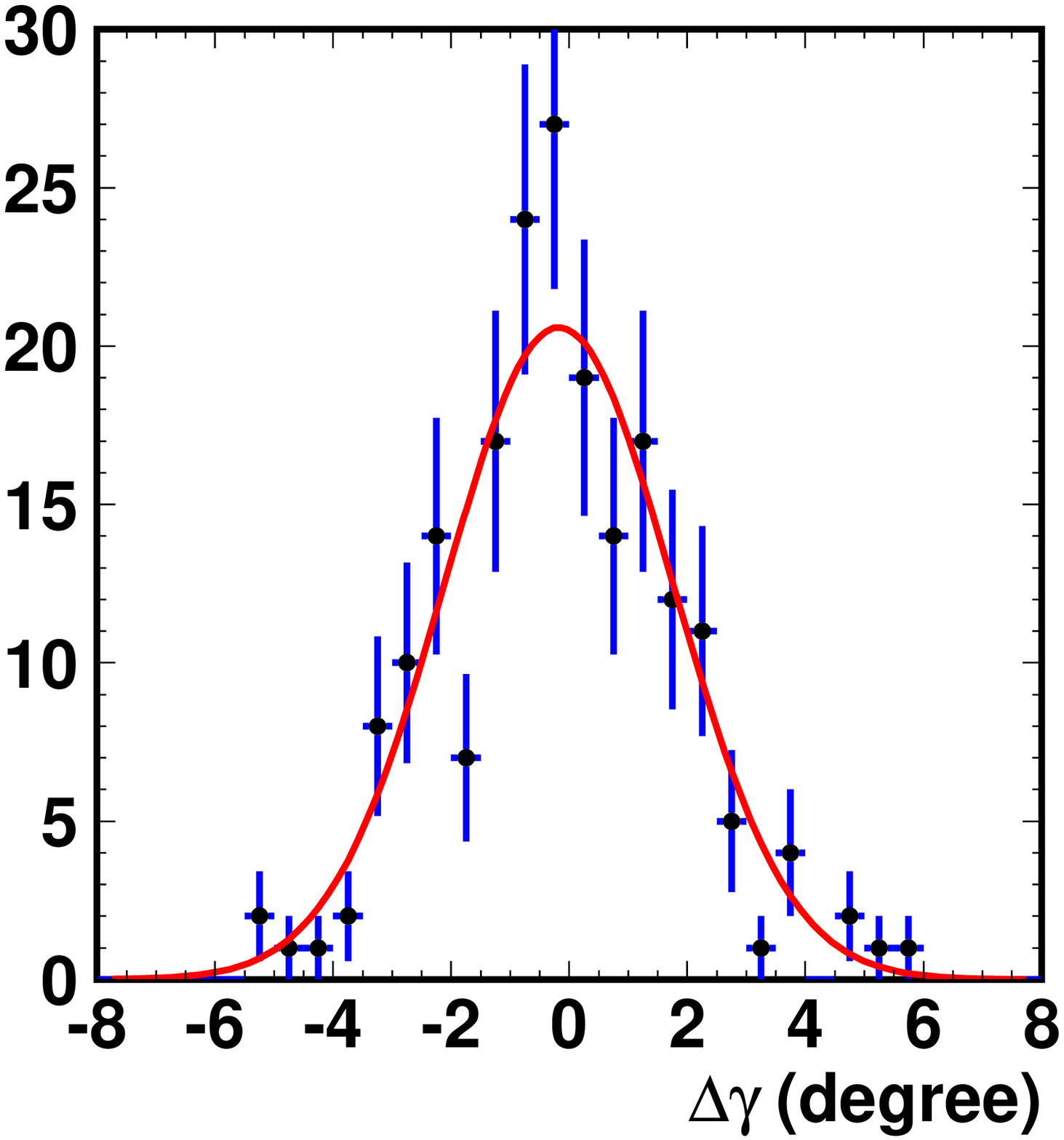, width=0.3\textwidth}
  \epsfig{file=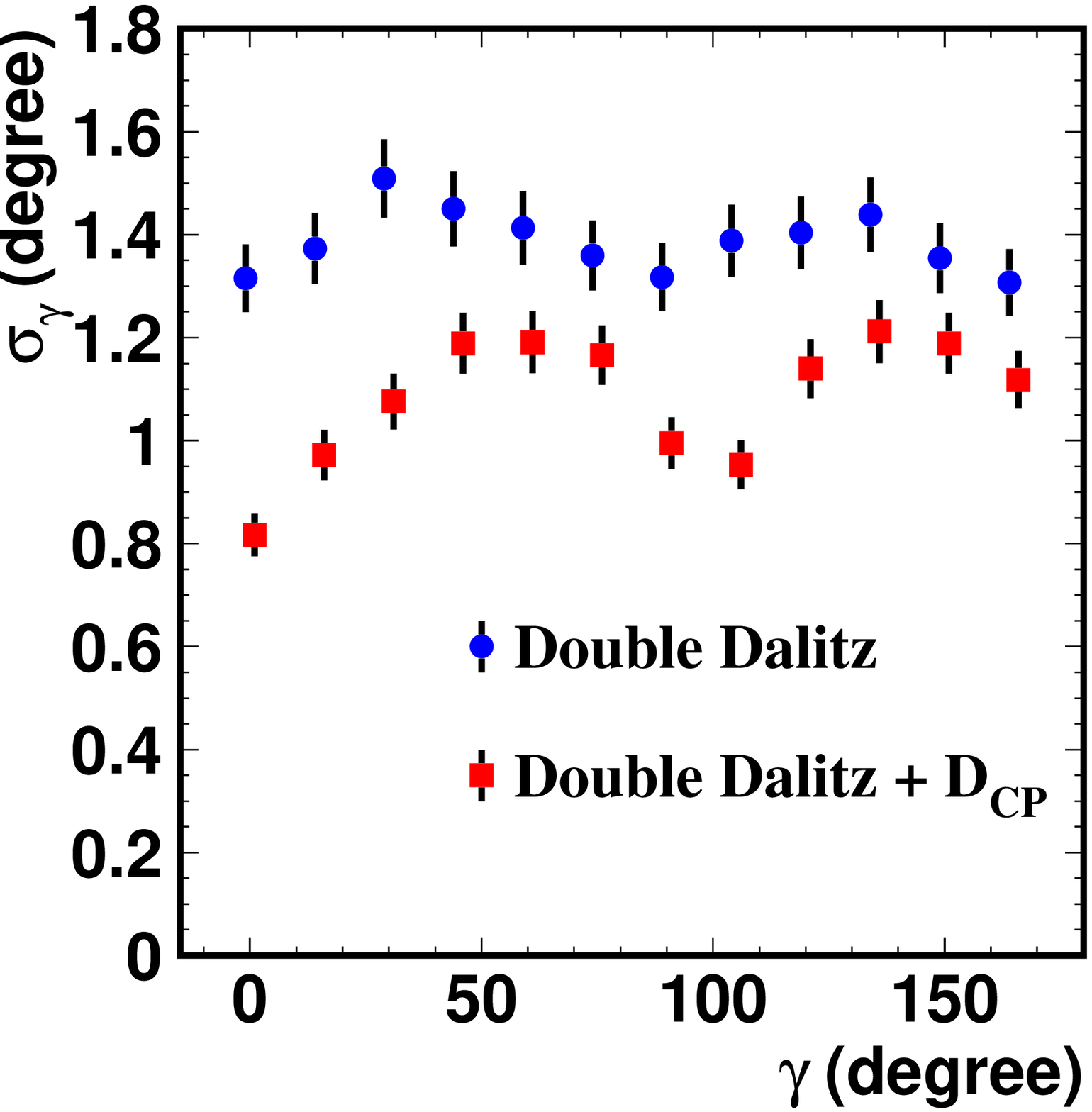, width=0.3\textwidth}
  \put(-340,130){\mbox{\bf(a)}}
  \put(-185,130){\mbox{\bf(b)}}
  \put(-30,130){\mbox{\bf(c)}}
  \caption{Accuracy of $\gamma$ measurement for the 
       binned double Dalitz fit of \bdkp, \dkpp (option 6)
       using model C. (a) Residual distribution for the 
       optimal binning, (b) residual distribution for 
       the rectangular binning with $Q=0.60$, (c) dependence of the 
       accuracy on the value of $\gamma$ and comparison 
       with the combined fit with \bdcpkp decays (option 8).
        }
  \label{dbldlz_resid}
\end{figure*}

An example of the residual distribution for $\gamma$ using option 6 (double
Dalitz plot fit) with amplitude model C is shown in 
Figure~\ref{dbldlz_resid}(a).  
No systematic bias is seen in any of the fits. 

We emphasise that, although the fit procedure uses a binning that depends 
on the decay amplitude, the method does not generate systematic bias even if 
the amplitude used to define the binning is wrong or the model-blind 
binning is used. To demonstrate this, we perform the fit of the data 
generated using model C, but using model-blind rectangular Dalitz plot 
bins 
(the bin boundaries in $m^2_{K\pi}$ are $M^2(K^*(892))$ and 
7~MeV$^2$/$c^4$; 
those in $m^2_{D\pi}$ are $M^2(D^*_2(2460))$ and 12~MeV$^2$/$c^2$, 
the bins above $m^2_{K\pi}=7$~MeV$^2$/$c^4$ are combined so that the 
number of bins is 7 as in the other fits). The quality factor 
for this binning is $Q=0.60$. The residual distribution for $\gamma$
using option 6 (double Dalitz plot fit)
is shown in Figure~\ref{dbldlz_resid}(b). The fit is unbiased, and the 
statistical accuracy compared to the fit with binning based on 
the correct underlying model is roughly proportional 
to the ratio of the $Q$ values: the $\gamma$ resolution is 
$\sigma(\gamma)=1.93\pm 0.10^{\circ}$. 

The statistical accuracy of the method does not depend significantly 
on the value of $\gamma$. The precision on $\gamma$ as a function of
$\gamma$ is shown in Figure~\ref{dbldlz_resid}(c) for model C, 
options 6 (double Dalitz plot fit only) and 8 (double Dalitz plot fit
combined with \bdcpkp). We expect that the dependence on the strong 
phases in \bdkp decays, which were taken arbitrarily in our study, 
should also not be large. 

\section{Conclusion}
\label{sec:conclusion}

We have proposed a technique to obtain the angle $\gamma$ of the Unitarity
Triangle in a model-independent way by using the decay chain \bdkp, \dkpp. 
Our method is independent of model assumptions in both \bdkp and \dkpp
amplitudes. 
The proposed approach allows not only to measure $\gamma$ in this decay chain,
but also to use the constraints on the \bdkp amplitude obtained in this
measurement for a model-independent \bdcpkp Dalitz plot analysis.

The precision of the $\gamma$ measurement using our technique will depend
strongly on the structure of \bdkp amplitude, which is as-yet unknown. 
However, MC simulation results show that if interference effects between
\bdnkp and \bdbkp amplitudes are strong enough to allow a meaningful
model-dependent measurement of $\gamma$ using \bdcpkp 
decays~\cite{Gershon:2008pe}, and the number of reconstructed \bdkp, \dkpp
decays is comparable to the number of \bdcpkp decays, the model independent
technique proposed here should enable a significant improvement in the
sensitivity.

We are grateful to our colleagues from the LHCb experiment,
and would particularly like to thank Alex Bondar and Mike Williams 
for discussions. We would like to acknowledge support from
the University of Warwick Research Development Fund,
the Science and Technology Facilities Council (United Kingdom) and from 
the Ministry of Science and Education of the Russian Federation.

\end{document}